\documentclass[preprint,showpacs,preprintnumbers,amsmath,amssymb,floatfix]{revtex4}

\usepackage{graphicx}
\usepackage{dcolumn}
\usepackage{bm}
\usepackage{rotating}
\usepackage{epsfig}

\begin{document}

\title{Cobalt-related impurity centers in diamond: electronic properties
and hyperfine parameters}

\author{R. Larico$^{(1)}$, L. V. C. Assali$^{(1)}$, W. V. M. Machado$^{(1)}$,
and J. F. Justo$^{(2)}$}

\affiliation{$^{\rm (1)}$ Instituto de F\'{\i}sica,
Universidade de S\~ao Paulo,\\
CP 66318, CEP 05315-970, S\~ao Paulo, SP, Brazil \\
$^{\rm (2)}$ Escola Polit\'ecnica, Universidade de S\~ao Paulo,\\
CP 61548, CEP 05424-970, S\~ao Paulo, SP, Brazil}

%\date{\today}

\begin{abstract}
Cobalt-related impurity centers in diamond have been studied using
first principles calculations. We computed the
symmetry, formation and transition energies, and
hyperfine parameters of cobalt impurities in isolated configurations and in
complexes involving vacancies and nitrogen atoms. We found that
the Co impurity in a divacant site is energetically favorable and segregates
nitrogen atoms in its neighborhood. Our results were discussed
in the context of the recently observed Co-related electrically active centers
in synthetic diamond.
\end{abstract}


\maketitle

\section{Introduction}
\label{sec1}

Semiconducting materials featuring wide bandgaps have potential technological
applications, such as ultraviolet radiation sources \cite{koizumi} or
high-voltage switching devices \cite{balmer}. Among those materials, diamond
has also received considerable attention from a fundamental standpoint, due
to a combination of superior physical properties. It has the largest known
bulk modulus, high thermal conductivity and chemical inertness. Growing
diamond out of graphite has been achieved by the high pressure-high
temperature (HPHT) methods. In order to speed up the process and allow to get
macroscopic samples, 3d-transition metal alloys have been used as
solvent-catalysts. Those transition metals (TM) end up contaminating the
samples, generating electrically and optically active centers \cite{collins1}.
Nickel was the first transition metal impurity unambiguously identified in
synthetic diamond \cite{spear}, ever since, several nickel-related active
centers have been observed by electron paramagnetic resonance
(EPR) \cite{isoya1,isoya2,pereira} and optical \cite{nazare1} measurements.
Although cobalt has been the most widely used solvent-catalyst to grow
diamond \cite{sung}, cobalt-related defects could not be identified as easily
as the nickel-related ones. It was usually considered that either cobalt atoms
were incorporated in very low concentrations to allow a proper detection or
that the cobalt-related centers were electrically and/or optically inactive.
However, recent investigations have identified several cobalt-related active
centers in synthetic diamond \cite{lawson,twitchen,iakou,nadolinny3}.

The presence of cobalt impurities in diamond was first reported in the
literature by EPR measurements \cite{bagda}. The unique feature of the
hyperfine structure was associated to a nuclear spin I = 7/2, a trace of
$\,^{59}$Co nuclei, and the center would be formed by an isolated
interstitial cobalt in an octahedral distorted environment, in a doubly
positive charged state (Co$_{\rm i}^{2+}$) with a 3d$^7$ configuration.
More recently, a cobalt-related center, labeled O4 \cite{twitchen}, was
identified in HPHT diamond using EPR measurements. The proposed microscopic
model for this center, with a spin S=1/2 in a monoclinic (C$_{\rm 1h}$)
symmetry, is a cobalt atom in a semi-divacancy site
${\rm (C_3 {\it V} Co {\it V} C_3)}$ plus a nitrogen atom replacing one of
the Co six nearest neighboring carbon atoms, in the negative charge state
${\rm (C_3 {\it V} Co {\it V} C_2 N)^-}$ \cite{nadolinny3}. Although there were
no resolved nitrogen-related EPR lines, the nitrogen was proposed to pertain
to this center due to two reasons: the presence of nitrogen atoms in the
complexes could explain
the symmetry lowering of the ${\rm (C_3 {\it V} Co {\it V} C_3)}$ precursor
complex and the resulting center would be similar to those previously
identified NE centers, that were related to nickel-nitrogen
complexes \cite{nadolinny}. Two additional cobalt-related active centers have
been identified in diamond, labeled as NLO2 and NWO1 \cite{nadolinny3}.
The similarity in the EPR spectra of O4, NLO2, and NWO1, all of them with
spin S=1/2, indicates that all those centers should involve a cobalt atom in a
semi-divacancy site with different number of neighboring nitrogen atoms.
Therefore, the microscopic model for either NLO2 and NWO1 was
suggested as ${\rm (N C_2 {\it V} Co {\it V} C_2 N)^0}$ or
${\rm (C_3 {\it V} Co {\it V} C N_2)^0}$, with C$_{\rm 2h}$ or C$_{\rm 1h}$
symmetry, respectively. Table \ref{tab1} summarizes the available experimental
data on the Co-related electrically active centers in diamond.

In order to understand the nature of those cobalt-related centers, and
eventually control their activity, it is important to have a deep knowledge
on their configurations, how they interact with other defects and/or
impurities, and the respective microscopic processes of formation. Theoretical
models can provide important information on those issues, such as the role of
those defects in the electronic and optical properties of the material and how
nitrogen atoms interact with those centers. Here, we used first principles
total energy calculations to investigate the properties of
several cobalt-related centers. Our results on symmetry, spin, formation and
transition energies, and hyperfine parameters were compared to the properties
of active centers recently identified in diamond.

\section{Methodology}
\label{sec2}

Earlier theoretical investigations on transition metals in diamond have been
focused on nickel-related defect centers \cite{gerstmann,goss,larico,larico2},
only very recently cobalt-related centers have received some attention
\cite{johnston,johnston2,larico10}. Here, we carried an investigation on
the cobalt-related centers using the all-electron spin-polarized full-potential
linearized augmented plane wave (FP-LAPW) method, implemented in the WIEN
package \cite{blaha}. The calculations were performed within the framework of
the density functional theory and the generalized gradient approximation,
using the Perdew-Burke-Ernzerhof exchange-correlation potential \cite{pbe},
and includes spin-orbit coupling in a second-variational procedure.
All the calculations were performed considering a 54-atom reference supercell.
We used atomic spheres for all atoms with radius of R =  1.2 a.u., a value
that allowed simulations with large relaxations and distortions without any
atomic sphere overlap. We additionally considered a 2 $\times$ 2 $\times$ 2
k-point grid to sample the irreducible Brillouin zone, as well as the
$\Gamma$-point.

Convergence on the total energy was achieved using an augmented-plane-wave
basis to describe the interstitial region, with the number of plane-waves
limited by the 7.0/R parameter. Self-consistent interactions were performed
until total energy and total charge in the atomic spheres changed by less
than $10^{-4}$ Ry/atom and $10^{-5}$ electronic charges/atom between two
iterations, respectively. The atomic positions were relaxed until the forces
were smaller than 10$^{-3}$ Ry/a.u., with no symmetry constrains. Those
approximations and convergence criteria have provided an accurate description
of several defect centers in semiconductors \cite{assali,ayres,bn}.
Formation and transition energies of all centers were computed using
the procedure discussed in ref. \cite{assali}. This procedure
requires the total energies of the respective defect center and the
chemical potentials of carbon, nitrogen and cobalt. Those chemical potentials
were computed using the total energy of carbon in a diamond lattice,
nitrogen in a N$_2$ molecule, and cobalt in a hcp lattice, all
computed within the FP-LAPW methodology with those same convergence criteria.

The isotropic hyperfine fields ($A_{iso}$) were computed as a combination
of the contact term plus the isotropic part of the orbital term, since
that for transition metals the orbital contribution is not quenched
and cannot be neglected. %, as it has been pointed out in Ref. \cite{amm}.

\section{Electronic and Structural Properties}
\label{sec3}

\subsection{Isolated Cobalt}

We start considering the cases of an isolated substitutional cobalt impurity.
According to symmetry considerations, in a tetrahedral crystalline
environment, the 3d-related Co atomic orbitals split into a three-fold $t_2$
plus a two-fold {\it e} irreducible representations. Since Co interacts strongly
with the nearest neighboring carbon atoms, the $e$ states are located below
the $t_2$ ones. While the {\it e} levels remain non-bonding, the d-related
$t_2$ states interact with the bandgap vacancy-related $t_2$ ones, that came
from the dangling bonds on the carbon host atoms surrounding the vacant site
into which the impurity was introduced. The resulting structure comprises of
a bonding $t_2$ level, resonant in the valence band (VB), plus an anti-bonding
vacancy-like $t_2$ level in the bandgap. This electronic structure is similar
to that of a substitutional nickel impurity in diamond \cite{larico}, both
being well described by the vacancy model \cite{watkins}.

In the positive charge state (Co$_{\rm s}^+$), the center is stable in
T$_{\rm d}$ symmetry and presents a closed shell configuration (S = 0). There
are three cobalt-related energy levels: the non-bonding {\it e} and the
bonding $t_2$ levels are resonant in the VB, while the anti-bonding $t_2$ one
remains unoccupied in the bandgap \cite{larico10}. In the neutral charge
state, the center presents spin S = 1/2 in a near tetragonal symmetry, with
the unpaired electron occupying the anti-bonding $t_{2 \uparrow}$. In the
negative charge state, the center presents a spin S = 1 in a near tetragonal
symmetry. In the doubly negative charge state (Co$_{\rm s}^{2-}$), the center
has a T$_{\rm d}$ symmetry and a spin S = 3/2. Table \ref{tab2} summarizes all
those results.

The electronic structure of the Co$_{\rm s}^{2-}$ center is similar to that of
Ni$_{\rm s}^{-}$ (known as the W8 center in the literature) \cite{isoya1,larico},
having a $t_{2\uparrow}^3t_{2\downarrow}^0$ bandgap configuration. In that sense, the
Co$_{\rm s}^{2-}$ center should be identified by EPR measurements as easily as the
Ni-related W8 center has been \cite{isoya1}, but no EPR signal has been
associated to such a Co-related center so far. This apparent paradox
could be explained by analyzing the respective positions of the transition
energies with relation to the nitrogen donor transition, that lies at
$\varepsilon_{\rm v}+3.8$ eV, where $\varepsilon_{\rm v}$ is the VB top. The
Ni$_{\rm s}$ (0/-) transition is at $\varepsilon_{\rm v}+3.0$ eV, therefore
below the nitrogen donor level, but the Co$_{\rm s}$ (-/2-) one is at
$\varepsilon_{\rm v}+4.0$ eV, above that donor level. As a result, the doubly
negative charge state of Co$_{\rm s}$ is probably unaccessible. The electronic
structure of the Co$_{\rm s}^{-}$ center is similar to that of
Ni$_{\rm s}^{0}$ \cite{larico}, both with spin S = 1, but there is
no evidence, from EPR measurements, from either one. The symmetry
lowering-related splitting between the two highest occupied levels in
Co$_{\rm s}^-$ and Ni$_{\rm s}^{0}$ is very small and may explain why those
centers have not been observed so far, as discussed in
Refs. \cite{baker,baker2}.
The Co$_{\rm s}^0$, with a spin S = 1/2, could be experimentally observed by
boron codoping, in order to provide compensated samples.

According to table \ref{tab2}, the formation energies of isolated cobalt
in interstitial sites (Co$_{\rm i}$) are considerably larger than those in
substitutional ones. This result is typical for transition metal impurities
in semiconductors with small lattice parameters, such as diamond \cite{larico},
silicon carbide \cite{assali}, and boron nitride \cite{bn}. Although this
would suggest a prevailing concentration of substitutional TMs,
growth conditions of HPHT diamond could still lead to interstitial TM
impurities.

The 3d-related electronic levels of Co$_{\rm i}$ centers present a $t_2$ state
below the {\it e} one. This indicates that the interstitial cobalt interacts
more weakly with its nearest neighbors than with its next nearest neighbors,
suggesting that the Ludwig-Woodbury (LW) model is well suited to describe
these centers \cite{lw}. On the other hand, in all charge states, the 3d-related
levels occupations do not follow the Hund's rule and Co$_{\rm i}$ presents a
low spin (LS) ground state configuration, which follows from the prevailing
crystal field splitting over the exchange one. These results contrast with those
of TM impurities in silicon, in which both the LW model and the Hund's rule
are generally followed.

For all Co$_{\rm i}$ charge states, the $t_2$ levels are fully occupied, being
resonant in the VB. In the negative charge state, Co$_{\rm i}^-$ presents
T$_{\rm d}$ symmetry, being diamagnetic (S = 0). In the neutral charge state,
Co$_{\rm i}^0$ presents a spin S = 1/2 and tetragonal symmetry. In the positive
charge state, Co$_{\rm i}^+$ presents a T$_{\rm d}$ symmetry and a spin S = 1,
with the highest occupied energy level being an $e_{\uparrow}^2$ near the VB top.
In the doubly positive charge state, Co$_{\rm i}^{2+}$ presents a spin S = 3/2
and tetragonal symmetry. For the latest case, the Co impurity behaves as an
acceptor impurity, leaving a polarized hole in the valence band top. This
behavior could be explored in the context of high temperature ferromagnetism
mediated by free carriers in semiconductors \cite{dietl}.

Several Co-related active centers appear in HPHT diamond after thermal
annealing at T$\approx 1800^0$\,C. It has been suggested that the resulting centers
are aggregates of  Co impurities and vacancies, that became mobile under
such conditions \cite{nadolinny3}. Annealing under even higher temperatures
(T$\approx 2000^0$\,C) leads to diffusion of nitrogen impurities, which pair
with those Co-vacancy complexes.
Figure \ref{fig0}a presents the atomic configuration of a Co impurity between
two carbon vacancies, which is labeled as $\rm (C_3 {\it V} Co {\it V} C_3)$.
Atoms from 1 to 6 represent the six neighboring carbon atoms of the Co impurity.
They could be better described as two sets of carbons in a trigonal symmetry
(fig. \ref{fig0}b), each one next to a vacant site.

Table \ref{tab2} presents the results for isolated cobalt in a divacancy
site. Figure \ref{fig1}b shows the electronic structure of the
${\rm (C_3 {\it V} Co {\it V} C_3)}^0$ center. It can be described
as an interaction between the anti-bonding divacancy states \cite{assali20}
(Fig. \ref{fig1}a) and the Co-related atomic orbitals (Fig. \ref{fig1}c).
The center has a spin S = 3/2 and the highest occupied level has a prevailing
Co 3d-character. In the positive charge state, the center has a D$_{3d}$
symmetry and a spin S = 2. The highest occupied level $e_{u \uparrow}^2$ is
near the top of the VB and has a prevailing divacancy-like behavior,
which resembles the result for the ${\rm (C_3 {\it V} Ni {\it V} C_3)}$ complex
in diamond \cite{larico3}. In the negative charge state, this center has also a
D$_{3d}$ symmetry and spin S = 1, with the highest occupied level being a
cobalt-related orbital. In the doubly (triply) negative charge state, it has
a C$_{\rm 1h}$ (D$_{3d}$) symmetry and spin S = 1/2 (S = 0). Recently, the
${\rm (C_3 {\it V} Co {\it V}C_3)}^-$ center has been suggested as the
microscopic model to explain the EPR data in Co-doped diamond \cite{nadolinny3},
in the same way as the ${\rm (C_3 {\it V} Ni {\it V}C_3)^-}$ was suggested as
the structure of the NE4 center \cite{nadolinny}. However, our results indicate
that the electronic structure of the ${\rm (C_3 {\it V} Co {\it V}C_3)}^-$
center cannot be associated to a pure 3d-$t_2$ low spin configuration which
would come from the remaining 3d-electrons of the Co impurity after all the
divacancy dangling bonds were passivated, as discussed in Ref. \cite{nadolinny3}.

\subsection{Cobalt-Nitrogen complexes}

We now discuss the properties of cobalt-related centers in a semi-divacancy
site with several nearby substitutional nitrogen impurities. Table \ref{tab3}
presents the results for complexes involving one or two nitrogen atoms.
The $\rm (C_3 {\it V} Co {\it V} C_2 N)$ center is described by one of those
six neighboring carbon atoms being replaced by a nitrogen one in the
${\rm (C_3 {\it V} Co {\it V} C_3)}$ precursor. The $\rm (N C_2 {\it V} Co
{\it V} C_2 N)$ center is a configuration with two substitutional nitrogen
atoms (in positions 3 and 6 of figure \ref{fig0}) at opposite sides of the divacancy,
while in the $\rm (C_2 N {\it V} Co {\it V} N C_2)$ center, the nitrogen atoms
are in positions 1 and 3. Finally, the $\rm (C_3 {\it V} Co {\it V} C N_2)$ center
has two nitrogen atoms at positions 2 and 3. Table \ref{tab4} presents the
results for centers involving three and four nitrogen atoms.
The $\rm (C N_2 {\it V} Co {\it V} N C_2)$ center has nitrogen atoms in positions
2, 3 and 5, while in the $\rm (C_3 {\it V} Co {\it V} N_3)$ center, the nitrogen atoms
are in positions 2, 3 and 4. The $\rm (N C_2 {\it V} Co{\it V} C N_2)$ center has
nitrogen atoms in positions 1, 2 and 5. The $\rm (C N_2 {\it V} Co {\it V} N_2 C)$
center has nitrogen atoms in positions 3, 4, 5 and 6.

The ${\rm (C_3 {\it V} Co {\it V} C_2 N)}$ complex, at all charge states described
in table \ref{tab3}, presents fully occupied Co 3d-related levels, such that
the relevant magnetic, and probably optical, properties come from the partially
occupied divacancy-related levels. An ionic model has been used to describe the
electronic structure of the $\rm (C_3 {\it V} Co {\it V} C_2 N)$ centers,
suggesting that the nitrogen impurity would work as a donor, giving away one
electron to the ${\rm (C_3 {\it V} Co {\it V} C_3)}$ precursor \cite{nadolinny3}.
Although such a simple model could be invoked to explain the driving force for
the complex formation \cite{zhao}, it is unsuitable to explain its final stable
configuration. This can be understood, for example, by analyzing the electronic
structure of the ${\rm (C_3 {\it V} Co {\it V} C_2 N)^0}$ complex. Considering
an ionic model, this complex should have a spin S = 1, that would be obtained by
taking the results of the precursor complex (Fig. \ref{fig1}b), and simply
adding one electron to the system. Our results show otherwise, this center is
diamagnetic (S = 0), which is a result of a strong interaction among nitrogen,
cobalt and carbon crystalline orbitals. This indicates that the covalent interactions
play a fundamental role in stabilizing this complex, as it has already been
shown in a number of TM-related complexes in semiconductors \cite{assali98}.
When a nitrogen atom replaces one of the carbon atoms, arranged in an almost
octraedral environment around the Co-semi-divacancy complex (fig. \ref{fig0}),
the symmetry lowering splits the $e_u$-related divacancy energy levels into two
non-degenerated {\it a}+{\it a}$^\prime$ levels. Their covalent interaction with
the N$_{\rm s}$ orbitals, that introduces a non-degenerate level in the upper gap
region, pushes the energy level with an {\it a} representation toward the valence
band, leaving the other one ({\it a}$^\prime$) in the gap. This could be better
understood by comparing the electronic structures of the
${\rm (C_3 {\it V} Co {\it V} C_3 )^{-2}}$ and
${\rm (C_3 {\it V} Co {\it V} C_2 N)^-}$ complexes, that have the same number of
electrons, shown respectively in  figs. \ref{fig2}a and \ref{fig2}b.
The nitrogen incorporation in the precursor complex alters substantially
the final electronic structure of the center, i.e., the picture of a nitrogen
atom just donating electrons to that precursor complex is not valid.

Table \ref{tab3} also presents the results for Co-complexes involving two nitrogen
atoms. For all configurations considered here, the centers are diamagnetic in
the positive and negative charge states, while they have spin S=1/2 for the
neutral and doubly positive and negative charge states. Table \ref{tab4}
presents the results for Co-complexes involving three and four nitrogen
atoms. The trends on the electronic structure, as result of nitrogen
incorporation, is presented in figure \ref{fig2}. Going from a complex with
one nitrogen atom (Fig. \ref{fig2}b) to a complex with two nitrogen atoms
(Fig. \ref{fig2}c), the covalent interaction is strengthened, leading to a
deepening of the divacancy-related levels with {\it a} representation.
The 3d-related Co levels remain almost unaffected, with respect to the valence
band top, by the nitrogen incorporation. The same trend is observed for
additional nitrogen incorporation, as shown in Figs. \ref{fig2}d and \ref{fig2}e
for complexes with three and four nitrogen atoms, respectively. Based on the
formation energies of the Co-complexes presented in tables \ref{tab2},
\ref{tab3} and \ref{tab4}, we find that nitrogen incorporation into the
${\rm (C_3 {\it V} Co {\it V} C_3)}$ precursor is considerably favorable,
suggesting that nitrogen complexing is very likely in synthetic diamond.

\section{Summary}
\label{sec4}

In summary, we performed a theoretical investigation on the properties of
Co-related impurity complexes in diamond. We find that the formation energy of
Co in interstitial sites is considerably larger than that for Co in
substitutional or divacancy sites, suggesting a prevailing concentration of Co
in the later sites. We also show that the electronic structure of a Co
impurity in substitutional or divacancy sites can be well described in terms of a
vacancy model, and results from a hybridization between the vacancy-related
orbitals and the Co 3d-related ones. On the other hand, the electronic structure for
interstitial Co gives a low spin ground state configuration for all charge states.

These results could help into building microscopic models for the known
Co-related active centers in diamond \cite{bagda,twitchen,nadolinny3}.
The center identified in Ref. \cite{bagda} is fully consistent with
Co$_{\rm i}^{2+}$ structure in terms of symmetry (D$_{\rm 2d}$) and electronic
structure (spin S=3/2), but the isotropic hyperfine field in the Co nucleus
(216 MHz) is smaller than the experimental hyperfine parameters
observed for this center \cite{bagda}. We show that the earlier proposed
picture of this center, in terms of a pure 3d$^7$ configuration, is
misleading since there is a strong covalent interaction between the
impurity and its neighbors. Additionally, the large computed formation
energy for interstitial TM impurities suggests that other configurations
could explain this center. Although with different symmetries, we found several
centers with spin S=3/2, related to Co-substitutional and Co-divacancy complexes.

Some cobalt-related centers, O4, NLO2, and NWO1, have been identified with
very low symmetries \cite{twitchen,nadolinny3}, which led to interpretations
that such symmetries should come from nearby defect and/or impurities, such as
vacancies and nitrogen atoms, although the EPR lines could not resolve
the presence of nitrogen. Our results show that cobalt centers in isolated
configurations may lead to low symmetries, even without such nearby defects.
Therefore, in order to discuss the low symmetry cobalt-related centers,
isolated cobalt could not be ruled out. The cobalt-related  O4, NLO2, and NWO1
centers have spin S=1/2. Considering our results, they could be related to
Co$_{\rm s}^0$, but the computed isotropic hyperfine field ($A_{iso}$) is too
small to explain the experimental values.

We also found several divacancy-related Co complexes, shown in tables \ref{tab3}
and \ref{tab4}, that could explain the experimental results in terms of
symmetry, spin, and hyperfine fields. The ${\rm (C_3 {\it V} Co {\it V} C_2  N)^-}$
configuration was proposed to explain the O4 center \cite{nadolinny3}. Our
results indicate that this configuration is consistent with the experimental
data in terms of spin and symmetry, but the computed isotropic hyperfine field
in the Co nucleus is much smaller (22 MHz) than the experimental value measured
on the O4 center (197 MHz). On the other hand, the same configuration in
a positive charge state,
${\rm (C_3 {\it V} Co {\it V} C_2 N)^+}$, has a higher isotropic hyperfine
field in the Co nucleus (176 MHz), being fully consistent with experimental data.
However, there are other complexes, involving several nitrogen atoms, that are
also consistent with experimental data for the O4 center.

For the  NLO2 and NWO1 centers, our results are consistent with the proposed
microscopic complexes with two nitrogen atoms, respectively
${\rm (N C_2 {\it V} Co {\it V} C_2 N)}$ and ${\rm (C_3 {\it V} Co {\it V} C N_2)}$,
but for a charge state 2+ and not neutral, as suggested in ref. \cite{nadolinny3}.
We find configurations involving two, three, and four nitrogen atoms in
different charge states that would also be fully consistent with those
experimental data.

Finally, we should stress that for all the centers studied here, the magnitude
of the hyperfine parameters in nitrogen nuclei, $A_{iso} {\rm (^{14}N)}$,
is inversely proportional to that in the cobalt nuclei, $A_{iso} {\rm (^{59}Co)}$.
Therefore, for those active complexes with low $A_{iso} {\rm (^{59}Co)}$,
the EPR-related nitrogen hyperfine lines should be observable.

\section*{Acknowledgments}

The authors acknowledge support from Brazilian agency CNPq and the
computational facilities provided by the CENAPAD-SP.

\pagebreak

\begin{table}[h]
\caption{Experimental results for Co-related EPR active centers in diamond.
The table presents the symmetry, spin (S), hyperfine parameters
($A_i$, $i$=1, 2, 3), and the suggested
microscopic model. The isotropic hyperfine fields ($A_{iso}$) were computed
as an average of the measured $A_i$. Hyperfine parameters are given in MHz.}
\label{tab1} \vspace*{0.2cm}
\begin{center}
\begin{tabular}{lcccccccc}
\hline \hline {Center~~~~~~~}       & {~~~Sym.~~~} & {~~~S~~~} &
{~~~$A_1$~~~} & {~~~$A_2$~~~} & {~~~$A_3$~~~} & {~~~$A_{iso}$~~~} & & Model \\
\hline
Co$_{\rm i}^{2+}$ $^{(a)}$ & C$_{3v}$ or D$_{\rm 2d}$ & 3/2 &
245 & 260 & 260 & 255 & & Co$_{\rm i}^{2+}$\\
O4 $^{(b)}$ & C$_{\rm 1h}$ & 1/2 &
248 & 180 & 163 & 197     & & ~${\rm (C_3 {\it V} Co {\it V} C_2 N)^-}$\\
NLO2 $^{(c)}$ & C$_{\rm 1h}$ or C$_{\rm 2h}$ & 1/2  &
230.8 & 183.9 & 161.2 & 192 &{\Large $\lceil$}  &
    ${\rm (N C_2 {\it V} Co {\it V} C_2 N)^0}$  \\
NWO1 $^{(c)}$ & C$_{\rm 1h}$ or C$_{\rm 2h}$ & 1/2 &
248 & -- & 187.4 &  & {\Large $\lfloor$} &  ${\rm (C_3 {\it V} Co {\it V} C N_2)^0}$\\
\hline \hline
\end{tabular}
\end{center}
$^{(a)}$ Ref. \cite{bagda}; $^{(b)}$ Ref. \cite{twitchen};
$^{(c)}$ Ref. \cite{nadolinny3}.
\end{table}
\vspace{1cm}
\pagebreak

\begin{table}[h]
\caption{Results for isolated Co impurity centers in diamond: symmetry, spin (S),
formation (E$_{\rm f}$) and transition (E$_{\rm t}$) energies, and isotropic
hyperfine fields ($A_{iso}$) at the $^{59}$Co nuclei. Energies and hyperfine
fields are given in eV and MHz, respectively. $\epsilon_{\rm F}$ is the
Fermi energy and transition energies are given with respect to the
valence band top. Theoretical approximations and numerical truncations lead
to estimated errors of $\approx$ 0.2 eV and $\approx$ 30 MHz in the
energies and hyperfine fields, respectively.}
\label{tab2} \vspace*{0.2cm}
\begin{center}
\begin{tabular}{lccclc}
\hline \hline
~~~~~~~Center  &{~~~Sym.~~~}& {~~~~S~~~~} &{~~~~~~~E$_{\mathrm
f}$~~~~~~~~} & {~~~~~E$_{\mathrm t}$~~~~~~} & ~~~~$A_{iso}$~~~~ \\
\hline
Co$_{\rm s}^+$  & T$_{\rm d}$  & 0
& $3.2 + \epsilon_{\rm F}$ & 3.0 (+/0) & 0 \\
Co$_{\rm s}^0$  & D$_{\rm 2d}$ & 1/2 & 6.2         &    & 76 \\
Co$_{\rm s}^-$  & D$_{\rm 2d}$ &  1
& $9.8 -\epsilon_{\rm F}$ & 3.6 (0/-)  & 77 \\
Co$_{\rm s}^{2-}$& T$_{\rm d}$ & 3/2
& $13.8 - 2\epsilon_{\rm F}$ & 4.0 (-/2-) & 78 \\[0.3cm]
Co$_{\rm i}^{2+}$&  D$_{2d}$   & 3/2
& $14.0 + 2 \epsilon_{\rm F}$ & 1.1 (2+/+) & 216 \\
Co$_{\rm i}^+$  & T$_{\rm d}$  & 1
& $15.1 + \epsilon_{\rm F}$ & 1.3 (+/0) & 138 \\
Co$_{\rm i}^0$  &  D$_{\rm 2d}$    & 1/2 & 16.4 &  & 143 \\
Co$_{\rm i}^-$  & T$_{\rm d}$  & 0
& $18.1 -\epsilon_{\rm F}$ & 1.7 (0/-) & 0 \\[0.3cm]
%${\rm (C_3 {\it V} Co {\it V} C_3)^{2+}}$ & C$_{1}$    & 3/2    &
%$5.3 + 2 \epsilon_{\rm F}$ &  0.81 & 71  & \\
${\rm (C_3 {\it V} Co {\it V} C_3)^{+}}$ &  D$_{3\rm d}$     & 2    &
$5.5 + \epsilon_{\rm F}$ & 0.3 (+/0) & 110 \\
${\rm (C_3 {\it V} Co {\it V} C_3)^{0}}$ &  C$_{\rm 2h}$     & 3/2  &
5.8 &                        & 108  \\
${\rm (C_3 {\it V} Co {\it V} C_3)^{-}}$ & D$_{3 \rm d}$     & 1    &
$6.5 - \epsilon_{\rm F}$ & 0.7 (0/-) & 43 \\
${\rm (C_3 {\it V} Co {\it V} C_3)^{2-}}$  & C$_{\rm 1h}$     & 1/2  &
 $7.4 - 2 \epsilon_{\rm F}$ & 0.9 (-/2-) & 106 \\
${\rm (C_3 {\it V} Co {\it V} C_3)^{3-}}$ & D$_{3 \rm d}$    & 0    &
$9.0 - 3 \epsilon_{\rm F}$ & 1.6 (2-/3-) & 0  \\
\hline \hline
\end{tabular}
\end{center}
\end{table}
\pagebreak

\begin{table}[h]
\caption{Results for Co-related defect centers in diamond involving
one and two nitrogen atoms: symmetry, spin, formation and transition
energies, and isotropic hyperfine fields at the $^{59}$Co nuclei.}
\label{tab3} %\vspace*{0.2cm}
\begin{center}
\begin{tabular}{lccclc}
\hline \hline ~~~~~~Center  &{~Sym.~}& {~~~~S~~~~}
&{~~~~~~E$_{\mathrm f}$~~~~~~~} & {~~~~~E$_{\mathrm t}$~~~~~~} & ~~~~$A_{iso}$~~~~ \\
\hline
${\rm (C_3 {\it V} Co {\it V} C_2 N)^{+}}$   & C$_{1 \rm h}$& 1/2   &
$3.9 + \epsilon_{\rm F}$ & 0.8 (+/0) & -176 \\
${\rm (C_3 {\it V} Co {\it V} C_2 N)^{0}}$  & C$_{1 \rm h}$  & 0  &
4.7 & & 0  \\
${\rm (C_3 {\it V} Co {\it V} C_2 N)^{-}}$  & C$_{1 \rm h}$  & 1/2  &
$5.7 - \epsilon_{\rm F}$ & 1.0 (0/-) & 22 \\
${\rm (C_3 {\it V} Co {\it V} C_2 N)^{2-}}$  & C$_{1 \rm h}$   & 0  &
$7.3 - 2 \epsilon_{\rm F}$ & 1.6 (-/2-) & 0 \\ [0.2cm]
${\rm (N C_2  {\it V} Co {\it V} C_2 N)^{2+}}$   & C$_{2 \rm h}$  & 1/2  &
$2.4 + 2 \epsilon_{\rm F}$ &  0.8 (2+/+) & 216 \\
${\rm (N C_2 {\it V} Co {\it V} C_2 N)^{+}}$ & C$_{2 \rm h}$  & 0 &
$3.2 + \epsilon_{\rm F}$ & 1.0 (+/0) & 0 \\
${\rm (N C_2  {\it V} Co {\it V} C_2 N)^{0}}$ & C$_{2 \rm h}$ & 1/2  &
4.2 &  & 17 \\
${\rm (N C_2  {\it V} Co {\it V} C_2 N)^{-}}$ & C$_{2 \rm h}$  & 0  &
$5.7 - \epsilon_{\rm F}$ & 1.5 (0/-) &  0 \\
${\rm (N C_2  {\it V} Co {\it V} C_2 N)^{2-}}$ & C$_{2 \rm h}$& 1/2 &
$8.9 - 2 \epsilon_{\rm F}$ & 3.2 (-/2-) & 17 \\[0.2cm]
${\rm (C_2 N {\it V} Co {\it V} N C_2)^{2+}}$ & C$_2$ & 1/2 &
$ 2.5  + 2 \epsilon_{\rm F}$ & 0.5 (2+/+) & -251 \\
${\rm (C_2 N {\it V} Co {\it V} N C_2)^{+}}$  & C$_2 $ & 0 &
$3.0 + \epsilon_{\rm F}$ & 0.9 (+/0) & 0 \\
${\rm (C_2 N {\it V} Co {\it V} N C_2)^{0}}$ & C$_2 $ & 1/2  &
3.9 & & 139 \\
${\rm (C_2 N {\it V} Co {\it V} N C_2)^{-}}$  & C$_2 $ & 0 &
$5.2 - \epsilon_{\rm F}$ & 1.3 (0/-) & 0 \\
${\rm (C_2 N {\it V} Co {\it V} N C_2)^{2-}}$ & C$_2 $ & 1/2 &
$9.2 - 2 \epsilon_{\rm F}$ & 4.0 (-/2-) & 8 \\[0.2cm]
%\hline
${\rm (C_3 {\it V} Co {\it V} C N_2)^{2+}}$ & C$_{1 \rm h}$ & 1/2 &
$ 2.2  + 2 \epsilon_{\rm F}$ & 0.5 (2+/+) & 153 \\
${\rm (C_3 {\it V} Co {\it V} C N_2)^{+}}$ & C$_{1 \rm h}$ & 0 &
$2.7 + \epsilon_{\rm F}$ & 0.9 (+/0)  & 0  \\
${\rm (C_3 {\it V} Co {\it V} C N_2)^{0}}$ & C$_{1 \rm h}$ & 1/2  &
$ 3.6$ & & -95 \\
${\rm (C_3 {\it V} Co {\it V} C N_2)^{-}}$ & C$_{1 \rm h}$ & 0 &
$ 5.0 - \epsilon_{\rm F}$ & 1.4 (0/-) & 0 \\
${\rm (C_3 {\it V} Co {\it V} C N_2)^{2-}}$ & C$_{1 \rm h}$ & 1/2 &
$8.8 - 2 \epsilon_{\rm F}$ &  3.8 (-/2-) & 6 \\
\hline \hline
\end{tabular}
\end{center}
\end{table}
%\vspace{1cm}

\pagebreak

\begin{table}[h]
\caption{Results for Co-related defect centers in diamond involving
three and four nitrogen atoms: symmetry, spin, formation and transition
energies, and  isotropic hyperfine fields at the $^{59}$Co nuclei.}
%\vspace*{0.2cm}
\begin{center}
\label{tab4}
\begin{tabular}{lccclc}
\hline \hline Center  &{~Sym.~}& {~~~~S~~~~} &
{~~~~~~E$_{\mathrm f}$~~~~~~~} & {~~~~~E$_{\mathrm t}$~~~~~~}  & ~~~~$A_{iso}$~~~~ \\
\hline
${\rm (C N_2 {\it V} Co {\it V} N C_2)^{+}}$ & C$_{\rm 1h}$  & 1/2  &
$1.9 + \epsilon_{\rm F}$ &  0.9 (+/0) & 274 \\
${\rm (C N_2 {\it V} Co {\it V} N C_2)^{0}}$ & C$_{\rm 1h}$  & 0    &
$2.8$ &  & 0 \\
${\rm (C N_2 {\it V} Co {\it V} N C_2)^{-}}$ & C$_{\rm 1h}$  & 1/2 &
$6.8- \epsilon_{\rm F}$ &  4.0 (0/-) & 9 \\[0.2cm]
${\rm (C_3 {\it V} Co {\it V} N_3)^{ +}}$ & C$_{\rm 1h}$& 1/2  &
$1.4 + \epsilon_{\rm F}$ &  1.0 (+/0) & 257 \\
${\rm (C_3 {\it V} Co {\it V} N_3)^{ 0}}$ & C$_{3v}$    & 0    &
$2.4 $ &  & 0 \\
${\rm (C_3 {\it V} Co {\it V} N_3)^{-}}$  & C$_{3v}$    & 1/2  &
$6.4 - \epsilon_{\rm F}$ &  4.0 (0/-) & 9 \\[0.2cm]
${\rm (N C_2 {\it V} Co {\it V}C N_2)^{2+}}$  & C$_1$    & 0    &
$1.0 + 2\epsilon_{\rm F}$  & 0.9 (2+/+) &  0 \\
${\rm (N C_2 {\it V} Co {\it V}C N_2)^{ +}}$  & C$_1$    & 1/2  &
$1.9 + \epsilon_{\rm F}$ & 1.3 (+/0) & 29 \\
${\rm (N C_2 {\it V} Co {\it V}C N_2)^{ 0}}$  & C$_1$    & 0    &
$3.2 $ & & 0 \\
${\rm (N C_2 {\it V} Co {\it V}C N_2)^{ -}}$  & C$_1$    & 1/2  &
$7.4 - \epsilon_{\rm F}$ & 4.2 (0/-) & 82 \\[0.2cm]
${\rm (C N_2 {\it V} Co {\it V} N_2 C)^{2+}}$ &C$_{\rm 2h}$ & 1/2    &
$0.1 + 2\epsilon_{\rm F}$ &  1.0 (2+/+) & 13 \\
${\rm (C N_2 {\it V} Co {\it V} N_2 C)^{+}}$  &C$_{\rm 2h}$ & 0    &
$1.1 + \epsilon_{\rm F}$ &  3.6 (+/0) & 0    \\
${\rm (C N_2 {\it V} Co {\it V} N_2 C)^{0}}$  &C$_{\rm 2h}$ & 1/2  &
$4.7$         &           &  210  \\
${\rm (C N_2 {\it V} Co {\it V} N_2 C)^{-}}$  &C$_{\rm 2h}$ & 1    &
$8.9 - \epsilon_{\rm F}$   & 4.2 (0/-) &  202  \\
\hline \hline
\end{tabular}
\end{center}
\end{table}

\pagebreak

\begin{figure}[h]
\caption{(a) Microscopic configuration of a Co impurity between two
carbon vacancies $\rm (C_3 {\it V} Co {\it V} C_3)$. The six nearest
neighboring carbon atoms are located in the edges of the gray triangles.
One set (1, 5, and 6 atoms) is next to one of the vacant sites while
the other (2, 3, and 4 atoms) is next to the other vacant site.
(b) Schematic representation of the configuration in a (111) plane.}
\label{fig0}
\end{figure}

\begin{figure}[h]
\caption{The calculated  electronic structure of the
${\rm (C_3 {\it V} Co {\it V} C_3)^0}$ center (b). This structure comes
from an interaction between the neutral divacancy levels (in D$_{3d}$
symmetry) (a) and the Co atomic states (d = 2$e_g$+$a_{1g}$) (c). Levels
with spin up (down) are represented by $\uparrow$ ($\downarrow$) arrows.
The number of filled (open) circles represent the electronic (hole) occupation
of each level.}
\label{fig1}
\end{figure}

\begin{figure}[h]
\caption{Electronic structure of the Co-related centers:
(a) ${\rm (C_3 {\it V} Co {\it V} C_3)^{2-}}$,
(b) ${\rm (C_3 {\it V} Co {\it V} C_2 N)^{-}}$,
(c) ${\rm (N C_2 {\it V} Co {\it V} C_2 N)^{0}}$,
(d) ${\rm (N C_2 {\it V} Co {\it V} C N_2)^{+}}$,
(e) ${\rm (C N_2 {\it V} Co {\it V} N_2 C)^{2+}}$,
all with spin S = 1/2. Levels with spin up (down) are represented
by $\uparrow$ ($\downarrow$) arrows. The number of filled (open)
circles represent the electronic (hole) occupation of each level.}
\label{fig2}
\end{figure}

\begin{figure}[h]
\includegraphics[width=140mm]{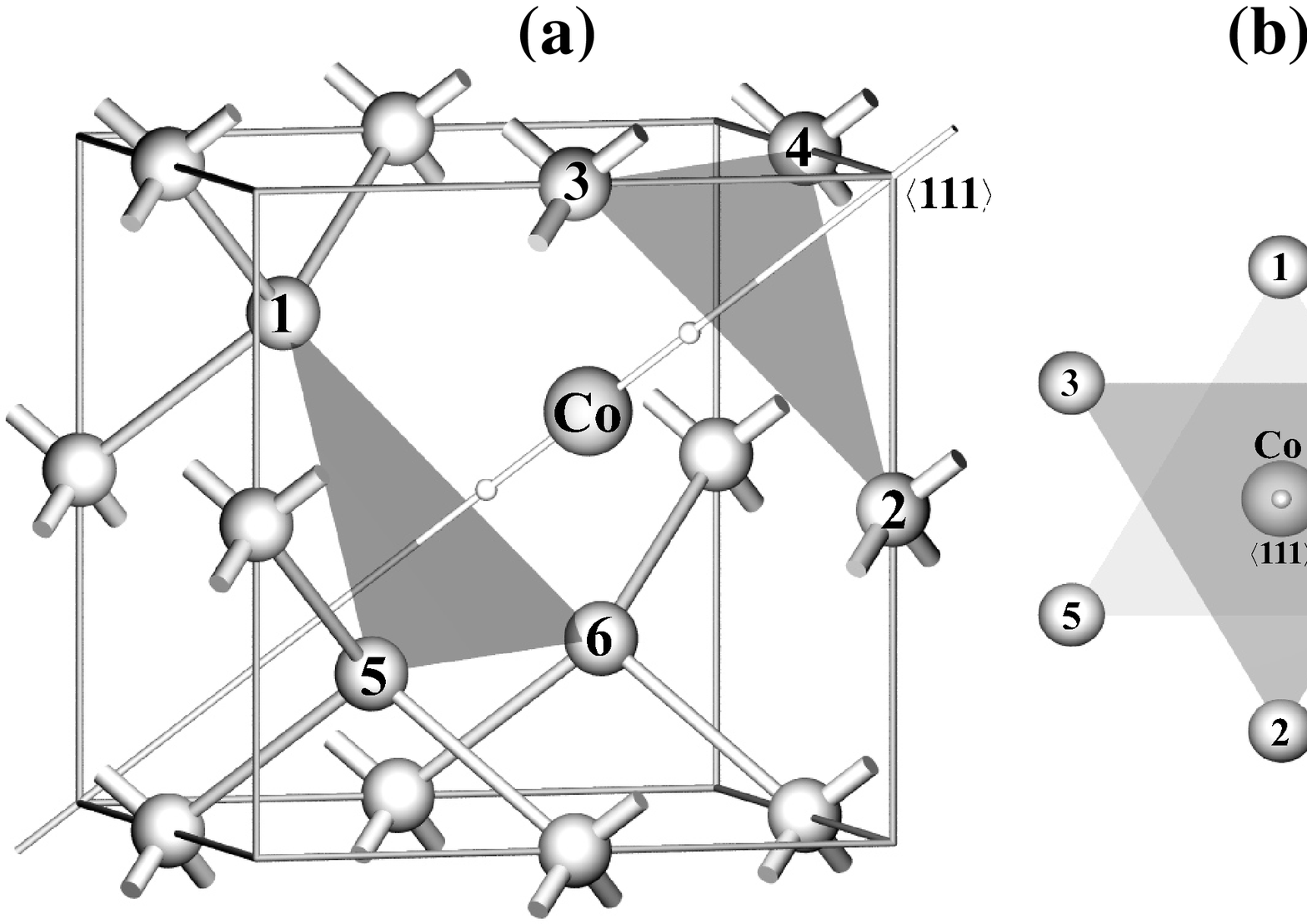}
\end{figure}

\begin{figure}[h]
\includegraphics[width=140mm]{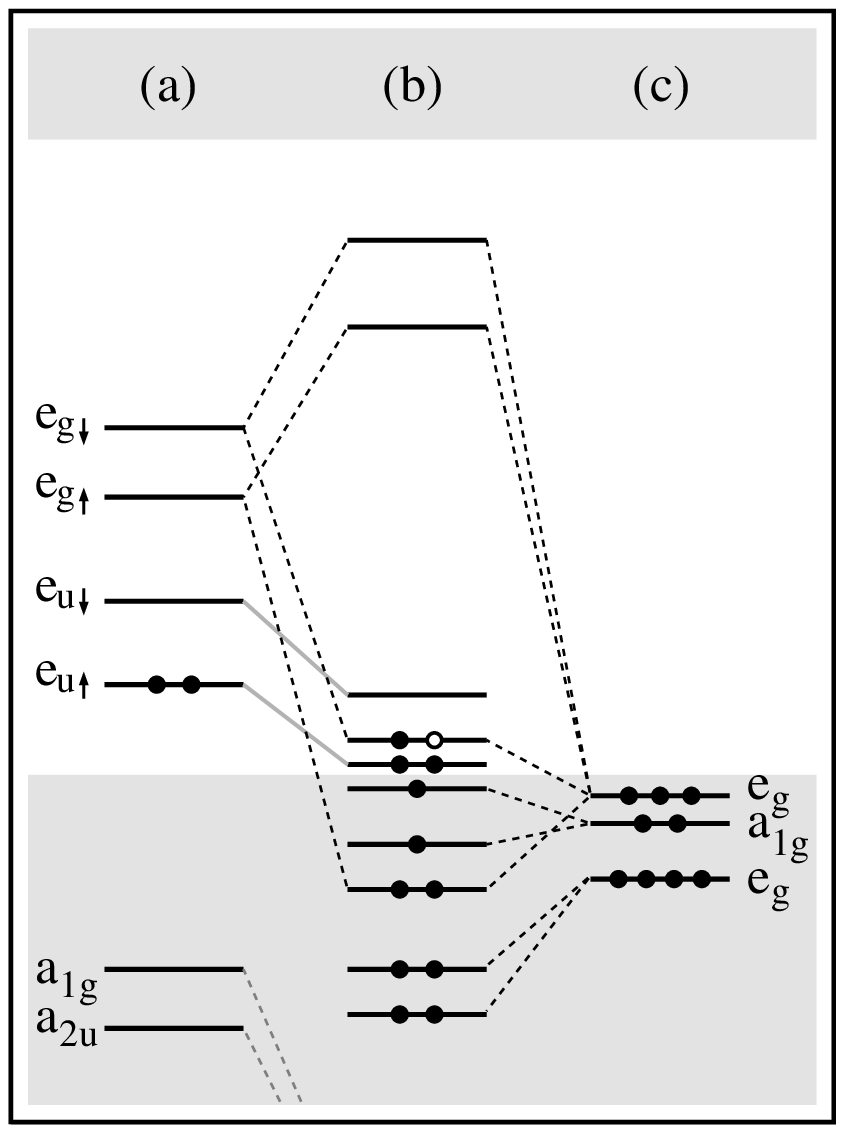}
\end{figure}

\begin{figure}[h]
\includegraphics[width=140mm]{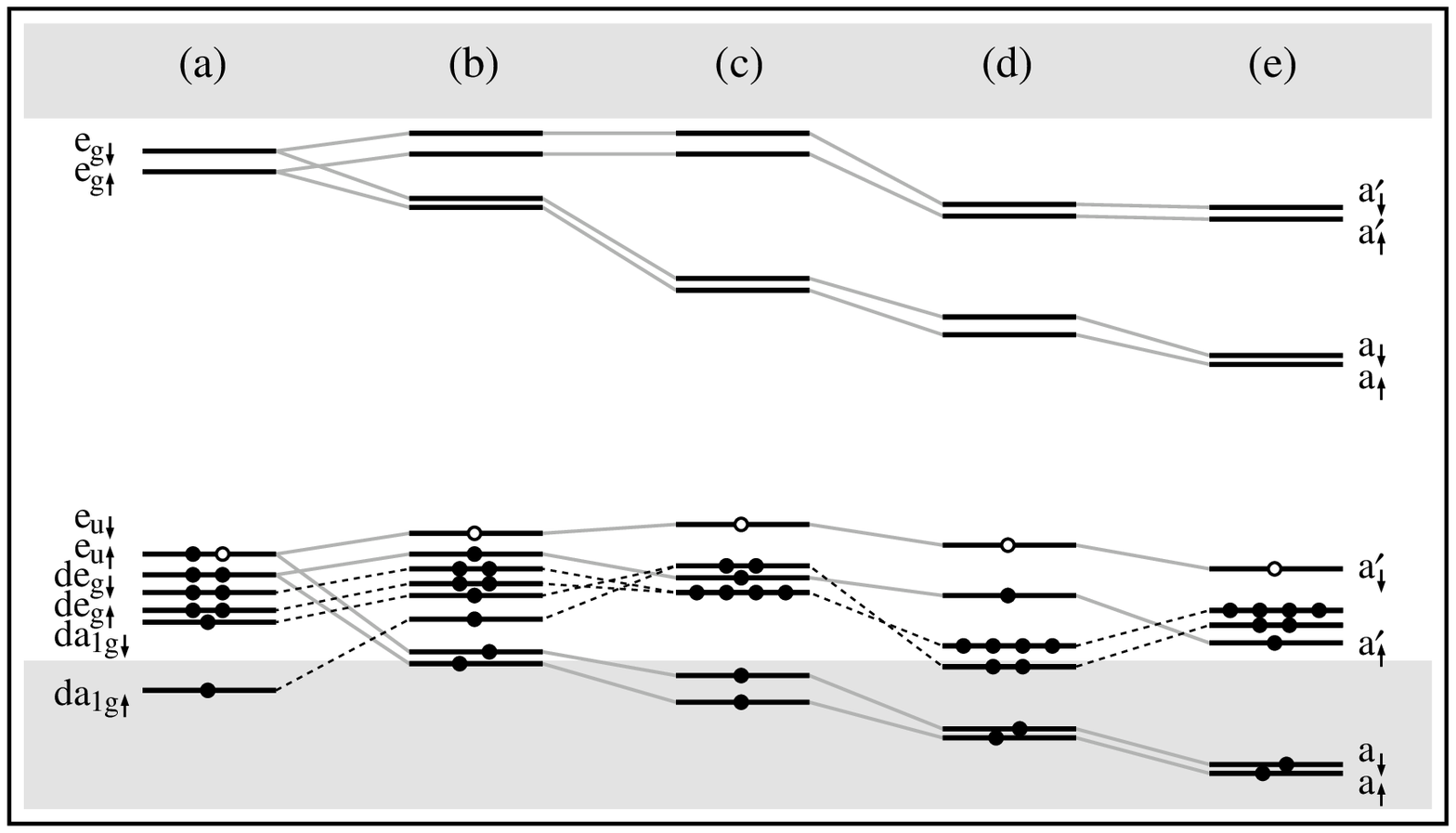}
\end{figure}

\end{document}